\documentclass[
 reprint,
 amsmath,amssymb,
 aps,superscriptaddress,
 longbibliography,
 longbibliography,
  prb,
 citeautoscript,
 citeautoscript,
twocolumn
]{revtex4-1}

\usepackage{graphics}
\usepackage{dcolumn}
\usepackage{bm}
\usepackage{amsmath}
\usepackage{epsfig}
\usepackage{braket}
\usepackage{commath}
\usepackage{color}
\usepackage[running]{lineno}

\usepackage{float} 

\usepackage[english]{babel}
\usepackage{color}

\newcommand{\beginsupplement}{%
        \setcounter{table}{0}
        \renewcommand{\thetable}{S\arabic{table}}%
        \setcounter{figure}{0}
        \renewcommand{\thefigure}{S\arabic{figure}}%
     }

\usepackage{blindtext}

\usepackage{color}

\def\ii{{\rm i}}  \def\EF{{E_{\rm F}}}  \def\dgr{d_{\rm gr}}
\def\chiexp{\chi_{\rm exp}^{(3)}}  \def\chisim{\chi_{\rm sim}^{(3)}}
  \def\chisim{\chi_{\rm sim}^{(3)}}

\usepackage{soul,xcolor} 
\setstcolor{red}

\begin{document}
\title{
Giant enhancement of third-harmonic generation in graphene-metal heterostructures}


\author{Irati Alonso Calafell*}
\affiliation{Vienna Center for Quantum Science and Technology (VCQ), Faculty of Physics, University of Vienna, Boltzmanngasse 5, Vienna A-1090, Austria}

\author{Lee A. Rozema}
\affiliation{Vienna Center for Quantum Science and Technology (VCQ), Faculty of Physics, University of Vienna, Boltzmanngasse 5, Vienna A-1090, Austria}

\author{David Alcaraz Iranzo}
\affiliation{ICFO-Institut de Ciencies Fotoniques, The Barcelona Institute of Science and Technology, 08860 Castelldefels (Barcelona), Spain}

\author{Alessandro Trenti}
\affiliation{Vienna Center for Quantum Science and Technology (VCQ), Faculty of Physics, University of Vienna, Boltzmanngasse 5, Vienna A-1090, Austria}

\author{Philipp K. Jenke}
\affiliation{Vienna Center for Quantum Science and Technology (VCQ), Faculty of Physics, University of Vienna, Boltzmanngasse 5, Vienna A-1090, Austria}

\author{Joel D. Cox}
\affiliation{Center for Nano Optics, University of Southern Denmark, Campusvej 55, DK-5230 Odense M, Denmark}
\affiliation{Danish Institute for Advanced Study, University of Southern Denmark, Campusvej 55, DK-5230 Odense M, Denmark}

\author{Avinash Kumar}
\affiliation{ICFO-Institut de Ciencies Fotoniques, The Barcelona Institute of Science and Technology, 08860 Castelldefels (Barcelona), Spain}

\author{Hlib Bieliaiev}
\affiliation{Vienna Center for Quantum Science and Technology (VCQ), Faculty of Physics, University of Vienna, Boltzmanngasse 5, Vienna A-1090, Austria}

\author{Sébastien Nanot}
\affiliation{ICFO-Institut de Ciencies Fotoniques, The Barcelona Institute of Science and Technology, 08860 Castelldefels (Barcelona), Spain}

\affiliation{Laboratoire Charles Coulomb (L2C), Université de Montpellier, CNRS, 34095 Montpellier Cedex, France}

\author{Cheng Peng}
\affiliation{Quantum Photonics Group, RLE, Massachusetts Institute of Technology, Cambridge, Massachusetts 02139, USA}
\author{Dmitri K. Efetov}
\affiliation{ICFO-Institut de Ciencies Fotoniques, The Barcelona Institute of Science and Technology, 08860 Castelldefels (Barcelona), Spain}
\author{Jin-Yong Hong}
\author{Jing Kong}
\author{Dirk R. Englund}
\affiliation{Quantum Photonics Group, RLE, Massachusetts Institute of Technology, Cambridge, Massachusetts 02139, USA}

\author{F. Javier Garc\'{i}a de Abajo}
\affiliation{ICFO-Institut de Ciencies Fotoniques, The Barcelona Institute of Science and Technology, 08860 Castelldefels (Barcelona), Spain}
\affiliation{ICREA-Institucio Catalana de Recerca i Estudis Avancats, Passeig Lluis Companys 23, 08010 Barcelona, Spain}

\author{Frank H. L. Koppens}
\affiliation{ICFO-Institut de Ciencies Fotoniques, The Barcelona Institute of Science and Technology, 08860 Castelldefels (Barcelona), Spain}

\author{Philip Walther}
\affiliation{Vienna Center for Quantum Science and Technology (VCQ), Faculty of Physics, University of Vienna, Boltzmanngasse 5, Vienna A-1090, Austria}

\date{\today}

\begin{abstract}
Nonlinear nanophotonics leverages engineered nanostructures to funnel light into small volumes and intensify nonlinear optical processes with spectral and spatial control. Due to its intrinsically large and electrically tunable nonlinear optical response, graphene is an especially promising nanomaterial for nonlinear optoelectronic applications. Here we report on exceptionally strong optical nonlinearities in graphene-insulator-metal heterostructures, demonstrating an enhancement by three orders of magnitude in the third-harmonic signal compared to bare graphene. Furthermore, by increasing the graphene Fermi energy through an external gate voltage, we find that graphene plasmons mediate the optical nonlinearity and modify the third-harmonic signal. Our findings show that graphene-insulator-metal is a promising heterostructure for optically-controlled and electrically-tunable nano-optoelectronic components.
\end{abstract}
\maketitle

The strong light-matter coupling regime can be reached by concentrating light into nanometric volumes, opening a wide range of applications that extend from optical sensing\cite{choi2011plasmonic} to quantum technologies\cite{tame2013quantum}.
To this end, metallic nanostructures that support plasmonic excitations - the coherent oscillations of conduction electrons - are widely used to intensify electromagnetic near fields, and turn out to be particularly important for enhancing nonlinear optical processes on the nanoscale \cite{Kauranen}. 
Although some aspects of plasmons can be modified by the geometry and optical properties of the host conductive media \cite{Stockman}, creating an actively tunable plasmonic platform remains an open challenge when relying on traditional plasmonic materials, such as noble metals \cite{chikkaraddy2016single,Abd,Echarri}. In addition, plasmons in noble metals face intrinsic ohmic losses that limit both their lifetimes and their optical nonlinearities \cite{maier2007plasmonics}.
This has motivated the development of schemes to mitigate such losses by, for example, employing lattice resonances in engineered nonlinear metasurfaces \cite{Kravets}.
Recently, a variety of nonlinear optical effects in graphene -- including third-harmonic generation (THG) \cite{Kumar,Jiang,Hong,Soavi}, four-wave mixing \cite{Hendry,Jiang}, the optical Kerr effect \cite{Dremetsika}, and high-harmonic generation \cite{Yoshikawa,Baudisch} -- have been observed, demonstrating that graphene exhibits an intrinsically strong and actively tunable nonlinear optical response.
Although, the optical nonlinearity of graphene is relatively efficient when normalised to the number of carbon atoms involved in the process, the atomically thin character of this material reduces the interaction volume. In this context, the use of photonic waveguides has been shown to substantially improve the nonlinear response of graphene \cite{Ooi}.

Besides possessing an intrinsically large optical nonlinearity, graphene can sustain surface plasmon polaritons.
These plasmons, which have exceptionally long lifetimes, are highly-confined and can be electrically tuned across a wide spectral range \cite{Ju, Koppens, Grigorenko, Garcia}.
Graphene-insulator-metal heterostructures have been used to demonstrate strong optical field confinement, down to single-atom length scales \cite{Alcaraz}, as well as near-perfect absorption of impinging light beams \cite{lee2019graphene,Thongrattanasiri,Kim}. It has also been argued that such systems can reach the strong-coupling quantum regime \cite{Gullans,Calafell,Koppens}.
Nonetheless, in spite of the extensive theoretical work predicting
coherent nonlinear plasmonic effects in graphene
 \cite{Mikhailov11,Gorbach,Gullans,Cox,Manzoni,Cox17,Rostami}, very few experiments have experimentally investigated plasmonic excitations in the nonlinear response regime of this material \cite{Constant,Kundys,Jadidi}. 
Moreover, the existing observations have required nonlinear mixing \cite{Constant}, THz radiation \cite{Jadidi}, or directly patterned graphene \cite{Kundys} to excite plasmons.

Here, we demonstrate efficient THG in graphene assisted by metallic elements that enhance light coupling into plasmons of the monatomic carbon layer. We study heterostructures made up of nanometer--thick gold nanoribbons placed above graphene, from which they are separated by an insulating spacer layer. Metal nanoribbons play a double role to intensify the electric field of a far-field mid-IR incident light beam into graphene and launch graphene surface plasmons \cite{Alcaraz}. 
Our experiments demonstrate that gold nanoribbons serve as efficient nanoantennas, increasing the observed THG intensity by three orders of magnitude above that of bare graphene.
We confirm that THG originates in the graphene layer, which allows us to actively tune the enhanced nonlinear signal by controlling the Fermi energy of graphene using an externally applied voltage.
This degree of tunability is simply not possible in conventional nobel metal plasmonics because the carrier density in those materials is too large to be substantially modified through gating.
Additionally, our experiments reveal the role of  graphene plasmons on the third-harmonic signal, emerging in particular witnessed when varying the carrier concentration.
We further observe that when the incident photon wavelength is tuned, the observed plamonic feature appears at different carrier concentrations, as predicted by our simulations, in which acoustic graphene plasmons are excited by the nanoribbons and confined below the metal.
Moreover, eliminating the plasmons from our model results in poor agreement with experimental data. These signatures of plasmon-enhanced and -suppressed third-harmonic generation provide a new route toward the amplification and control of light at deep subwavelength lengthscales.

\begin{figure*}[t]
\centering
\includegraphics[width=0.85\textwidth]{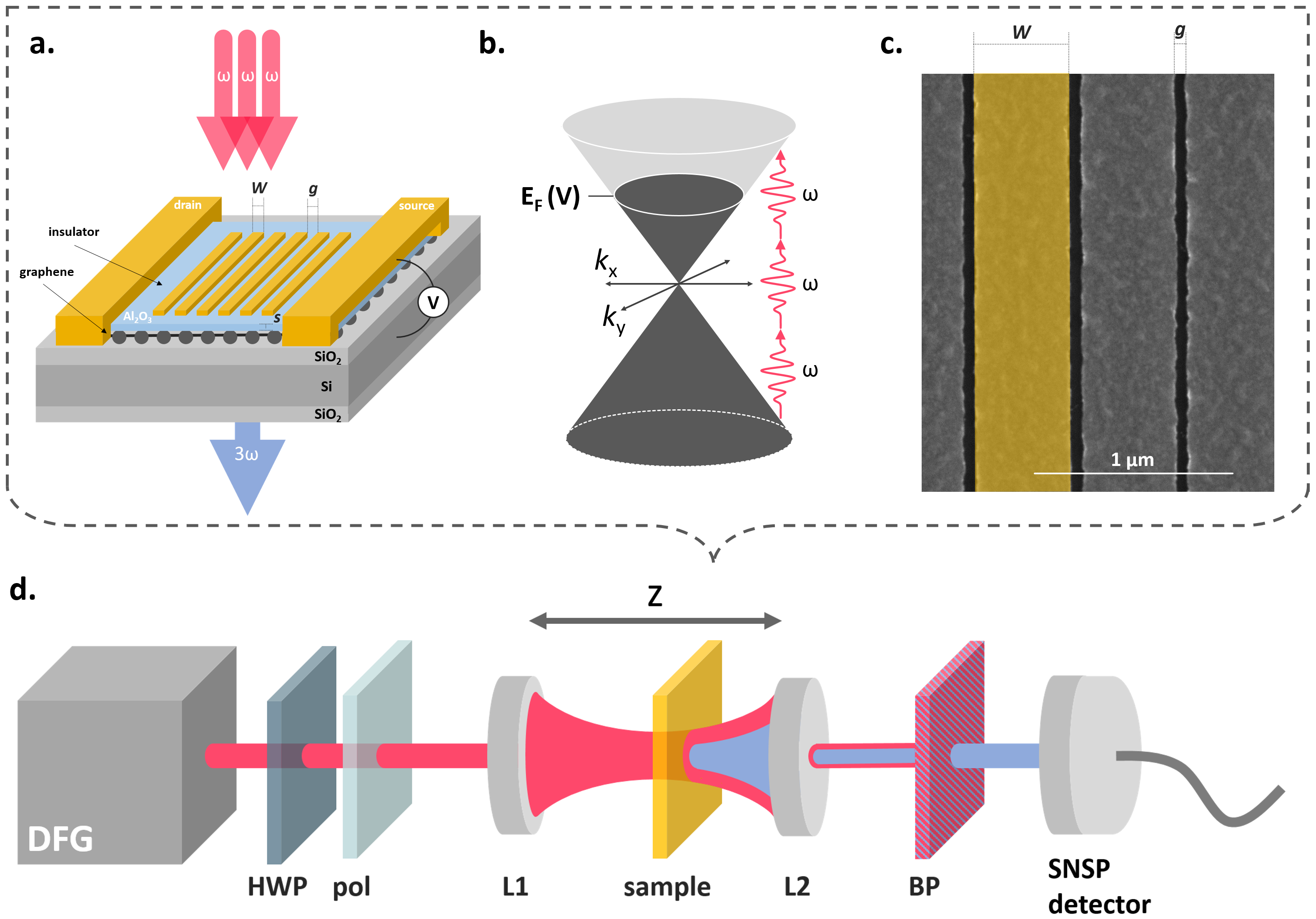}
\caption{\textbf{Gate-tunable graphene heterostructures.} \textbf{a.} Graphene is encapsulated by a few-nm-thick $\mathrm{Al_2 O_3}$ or a monolayer h-BN film, setting the space \textit{s} between the graphene and the gold nanoribbons. 
The gold nanoribbon arrays are characterised by the ribbon width \textit{W} and the inter-ribbon gap \textit{g}. Normally-incident light of frequency $\omega$ undergoes third-harmonic generation, which is collected in transmission. A gate voltage (V) tuned from $-150$\,V to $+150$\,V sets the graphene Fermi energy $\EF$.
\textbf{b.} The conical electron dispersion relation of graphene can be tuned in resonance with one, two, or three incident photons.
\textbf{c.} Scanning electron microscopy image of one of our high-quality gold nanoribbon arrays.
 \textbf{d.} Sketch of our experimental setup. Difference-frequency generation (DFG) between signal and idler beams of an optical parametric oscillator (not shown) provides mid-IR $\approx 260$\,fs pulses. A half-wave plate (HWP), together with a polariser (pol), selectively rotate the linear polarisation of the incident light, which is then focused onto the sample. A second lens collimates the incident and outgoing third harmonic light. A band-pass filter (BP) isolates the third-harmonic signal, which is coupled into a multimode fiber and sent to a superconducting-nanowire single-photon detector (SNSPD).
The sample is moved in the \textit{z} direction (z-scan), in and out of the focal point of the beam.
}
\label{img:Figure1}
\end{figure*}

\begin{figure*}[t]
\centering
\includegraphics[width=0.8\textwidth]{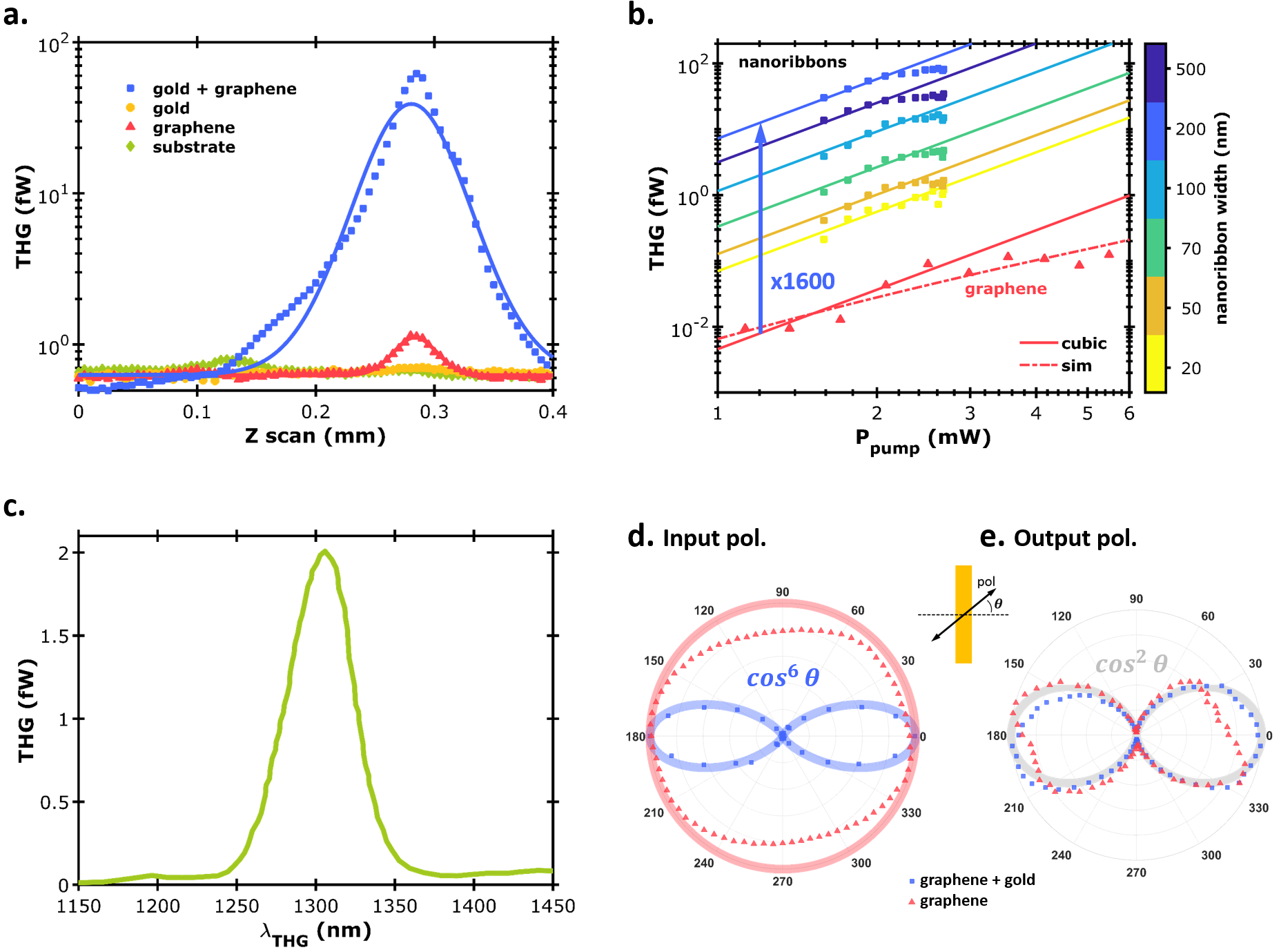}
\caption{\textbf{Characterization of the third-harmonic signal.} \textbf{a.} Z-scan measurements on four different regions of the sample (symbols). Gaussian fits to the data (curves) provide visual guides. 
\textbf{b.} Power scaling of the measured third-harmonic signal (symbols) in bare graphene and in gold-nanoribbon--graphene heterostructures for different nanoribbon widths with a monolayer h-BN spacer between the gold and graphene. The incident wavelength is $3.9\,\mathrm{\mu m}$ and the Fermi energy is $\EF\approx 150\,\mathrm{meV}$ for this measurement.
Linear fits to the data (solid lines) are used to determine the enhancement of the heterostructure relative to bare graphene. The dashed curve, which models the increase of the electron temperature with the incident light power, explains the observed saturation. 
\textbf{c.} Spectrum of third-harmonic generation from bare graphene, measured with $3.9\,\mathrm{\mu m}$ incident light.
\textbf{d.} Third-harmonic signal of bare graphene and heterostructures measured as the input polarisation is rotated. Bare graphene (red) is isotropic with respect to the incident polarisation, while the gold nanoribbons (blue) result in a $\cos^6{\theta}$ dependence on the polarisation angle $\theta$ relative to the direction perpendicular to the ribbons (\textit{i.e.}, as expected from a third-order power scaling). 
\textbf{e.} Third-harmonic emission measured for fixed input polarisation when a polariser placed after the sample is rotated. For bare graphene (red), the third-harmonic signal is co-polarised with the input light, while for the heterostructures (blue) the third-harmonic polarisation is always orthogonal to the nanoribbons.  In both cases, the polarisation of the third-harmonic signal is coherent relative to the incident light, which is indicated by the grey $\cos^2{\theta}$ line.
For the data presented in panels d and e, the incident light wavelength is $5.5\,\mathrm{\mu m}$ and the Fermi energy is $\EF\approx 150\,\mathrm{meV}$. The studied heterostructure has {an $\mathrm{Al_2 O_3}$ spacer of} $s = 5\,\mathrm{nm}$, and ribbons with $W = 200\,\mathrm{nm}$, and $g = 50\,\mathrm{nm}$.}
\label{img:Figure2}
\end{figure*}

\section*{THG in graphene heterostructures}
Our samples are van der Waals heterostructures, consisting of a graphene sheet with a metallic nanoribbon array placed a few nanometers above it and separated by {either} an insulating $\mathrm{Al_2 O_3}$ ($3-20\,\mathrm{nm}$) or a monolayer h-BN spacer, as depicted in Fig. \ref{img:Figure1}a.
All of the gold nanoribbons are $8\,\mathrm{nm}$ thick, with an additional $2\,\mathrm{nm}$ titanium adhesion layer between the spacer and the gold nanoribbons.
In order to isolate the nonlinear signal from the heterostructure, we use a modified z-scan setup with a tight depth of focus (Fig. \ref{img:Figure1}d). In our configuration, the sample is moved through the focal point of a fs-pulsed mid-IR incident light beam (with a wavelength of $5.5\,\mathrm{\mu m}$ nm or $3.9\,\mathrm{\mu m}$), and a third-harmonic signal (at $1.833\,\mathrm{\mu m}$ or $1.3\,\mathrm{\mu m}$, respectively) is measured in transmission (as detailed in the Methods section). All the measurements in this work are performed under ambient pressure and temperature conditions. A set of representative z-scan measurements is presented in Fig. \ref{img:Figure2}a, showing that we only observe signals from bare \textit{graphene}, and \textit{gold+graphene} heterostructures (see Figure S2 of the Appendix for more detail). The spectrum of the nonlinear signal (with $3.9\,\mathrm{\mu m}$ incident light) is presented in Fig. \ref{img:Figure2}c, showing a clear peak at the third-harmonic wavelength. The wavelength of the THG signal with $5.5\,\mathrm{\mu m}$ incident light is confirmed in Fig. S1 in the Appendix.

Our data clearly show that the THG signals from the heterostructures are greatly enhanced compared to bare graphene, and additional control experiments (see Figure S2 of the Appendix) show that the metal structures alone do not produce a measurable THG signal. Moreover, the THG signal is maximised when the polarisation is perpendicular to the direction of the nanoribbons (red squares in Fig. \ref{img:Figure2}d,e). Additionally, as shown in Fig. \ref{img:Figure2}d, the THG signal of bare graphene is co-polarised with the incoming light (red triangles), and the THG signal of the heterostructures is perpendicular to the nanoribbons (red squares). In both cases, the strongly polarised signal indicates a coherent nonlinear process.

\begin{figure*}[t]
\centering
\includegraphics[width=0.9\textwidth]{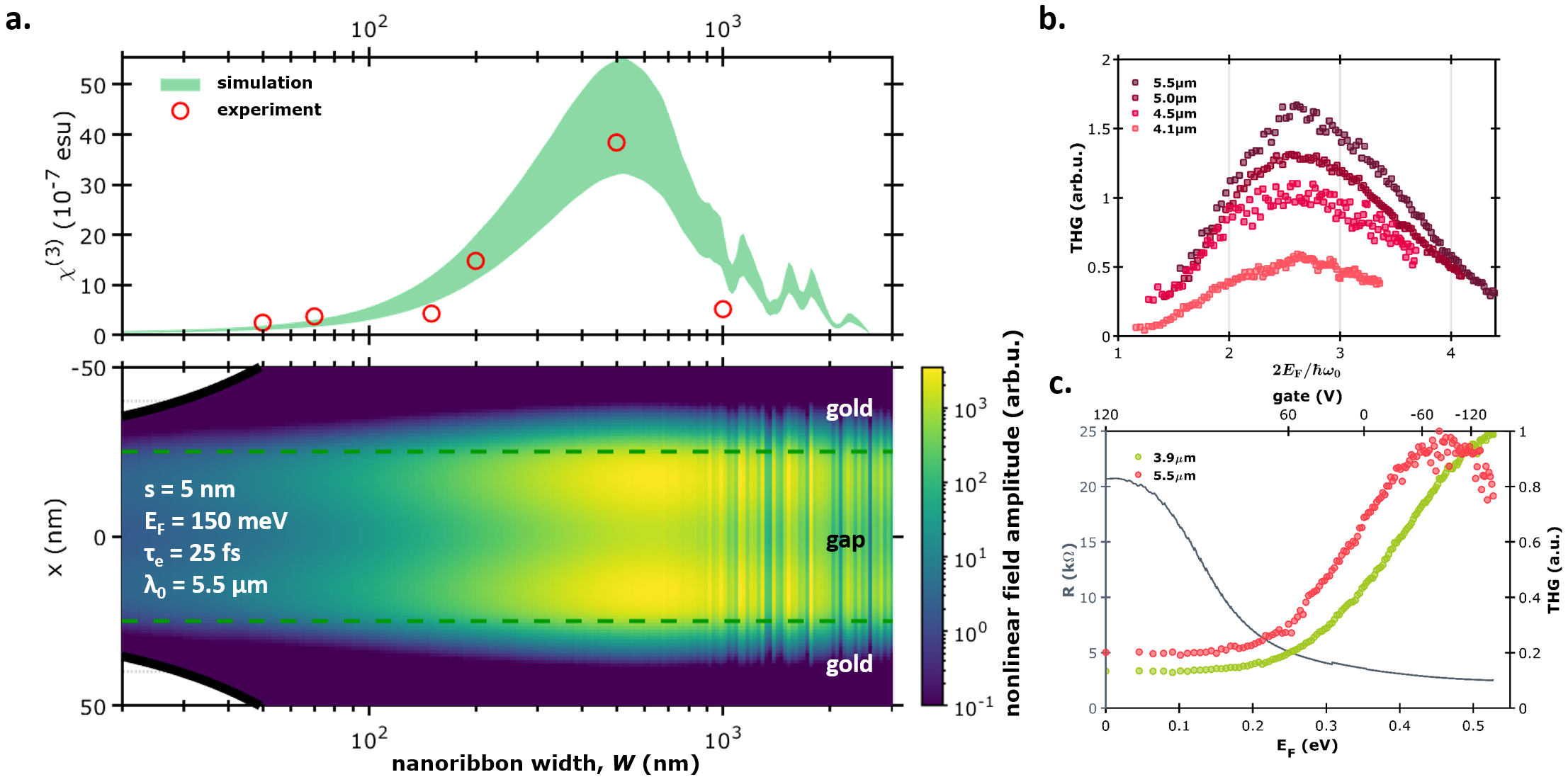}
\caption{\textbf{Third-order nonlinear susceptibility and electrical gating dependence.} 
\textbf{a.} Top: {Effective nonlinearity of gold-nanoribbon--graphene heterostructures versus ribbon width, with an $s = 5\,\mathrm{nm}$ $\mathrm{Al_2 O_3}$ spacer and an incident light wavelength of $5.5\,\mathrm{\mu m}$.
Symbols stand for experimentally estimated values, while the shaded area shows the result of our simulation in which an uncertainty of $20\%$ on the inter-ribbon gap is introduced to account for manufacturing imperfections.
The simulated $\chi^{(3)}$ is the result of integrating the nonlinear fields (plotted below) along $x$.} Bottom: Field confinement in the gap between the gold nanoribbons plotted along the gap (vertical axis) for different nanoribbon widths (horizontal axis). 
The colour plot indicates the calculated product of electric fields of third harmonic and incident light $E^3_{x(\omega)}E_{x(3\omega)}/E_{0}$, relevant for third-harmonic generation.
\textbf{b.} Gate-dependence of the third-harmonic signal in bare graphene for various incident light wavelengths.
\textbf{c.} Gate-dependence of the third-harmonic signal in heterostructures with $W=200\,\mathrm{nm}$, $g=50\,\mathrm{nm}$ and $s = 5\,\mathrm{nm}$, and incident light wavelengths of $3.9\,\mathrm{\mu m}$ and $5.5\,\mathrm{\mu m}$ (green and red symbols, respectively). The gray curve is the measured resistance of the the sample, the peak of which indicates the charge neutrality point.
}
\label{img:Figure3}
\end{figure*}

We quantify the enhancement and verify the third-order nature of our signal by measuring the power dependence of bare \textit{graphene} (red triangles) and the \textit{gold+graphene} heterostructures (squares), shown in Fig. \ref{img:Figure2}b. The slope of the linear fits on a log-log scale is fixed to $3$ and the \textit{y}-intercepts are free parameters, which we use to calculate the enhancement of the heterostructures over bare graphene. Although at higher powers a small saturation effect can be observed in the \textit{gold+graphene} data, a clear third-order power scaling is supported by the data.
To explain the saturation, we model the effect of the increasing incident light power increasing the electronic temperature (see Methods section \textit{Electron temperature}). The result for bare graphene, plotted as the dashed curve, fits our data well.
We experimentally find the maximum enhancement for a $3.9\,\mathrm{\mu m}$ incident wavelength with a monolayer h-BN spacer and a ribbon width of $W =200\,\mathrm{nm}$. 
Under these conditions, the heterostructures produce a THG signal that is  $1600\pm 800$ times larger than that of bare graphene, which corresponds to a maximum THG conversion efficiency of $2\cdot10^{-7}\,\%$ at $1.9\,\mathrm{mW}$ of incident power.

To understand the enhancement mechanism and the role of the metal, we perform rigorous coupled-wave analysis (RCWA) simulations that are presented in Fig. 3a, the details are presented in the Methods section. 
Notice that due to the form of the nonlinear field integral (Eq. \ref{eq:alpha_sim}), the analogous expression for higher-harmonics predicts an even larger enhancement.
The simulations show a strong concentration of the electric field in the gap between the nanoribbons only when the polarisation of the incident field is perpendicular to the ribbons. 
In contrast, the bare graphene signal is independent of the incident polarisation (red triangles Fig. \ref{img:Figure2}c).
Note that the slight asymmetry is caused by a polarisation-dependent detection efficiency of our superconducting detector. 
Therefore, we conclude that the enhancement is mediated by the gold nanoribbons, which amplify the electric near-field in the graphene layer. The simulations in Fig. \ref{img:Figure3}a also show that the field strength in the gap depends on the width of the nanoribbons. 
To verify this experimentally, we performed a series of THG measurements for different nanoribbon widths, with a spacer thickness of {$s = 5\,\mathrm{nm}$} and an incident light wavelength of $5.5\,\mathrm{\mu m}$ (results for other spacers are shown in Fig. S4). From the THG signals we estimate an effective third-order susceptibility $\chi^{(3)}$ as described in the Methods section. The result is shown in Fig 3a.  
The experimental data agree well with our {RCWA} simulations, which assume an uncertainty of $\pm 20\,\%$ on the gap size (nominally set to $50$\,nm) caused by experimental imperfections.

\section*{Electrical tunability}
While using gold nanribbons of the appropriate width can greatly enhance the nonlinear response of the system, this width cannot be actively changed.
In contrast, the optical nonlinearity in graphene can be electrically tuned, thus providing a practical route to active control.
The optical response in graphene depends on the strength of the intraband and interband transitions \cite{Mikhailov16,Soavi,Jiang,Rostami}, which in turn depends on the ratio of the impinging light energy $\hbar\omega_0$ to the graphene Fermi energy $\EF$. The latter can be tuned in-situ by applying an external voltage to the graphene layer relative to the silicon substrate.

Conceptually, we can understand the effect of $\EF$-tuning on THG as illustrated in Fig. 1b. By sweeping the gate voltage, one can match the Fermi energy to be resonant to an interband transition for one, two or three incident photons. Each transition results in a different resonance in the third-order nonlinear susceptibility given by,
\begin{equation}
\chi^{(3)}\propto\left[-17G\left(\frac{\hbar\omega_0}{2|\EF|}\right)+64G\left(\frac{2\hbar\omega_0}{2|\EF|}\right)-45G\left(\frac{3\hbar\omega_0}{2|\EF|}\right)\right]
\end{equation}
where $G(x)=\ln\left|\frac{1+x}{1-x}\right|+i\pi H(|x|-1)$ and $H$ is the Heaviside step function \cite{Cheng2015}.
However, the resonances enter the expression with different signs.
Thus, for small Fermi energies, $2\EF<\hbar\omega_0$, although the three transitions can occur, the total nonlinear susceptibility nearly cancels out.
For large Fermi energies, $2\EF>3\hbar\omega_0$, all three of these transitions are Pauli-blocked and there is only a non-resonant small intraband contribution. \cite{Jiang,Soavi}
At intermediate Fermi energies, $\hbar\omega_0<2\EF<3\hbar\omega_0$, it is possible to increase THG by, for example, Pauli blocking the one-photon and two-photon transitions, so that only the three-photon transition is allowed and the other two transitions no-longer cancel it out. For low electron temperatures the gate response is expected to result in several sharp features as the system is tuned in and out of resonance.  However, thermal broadening turns these features into broad shoulders \cite{Jiang,Soavi}.

\begin{figure*}[t]
\centering
\includegraphics[width=1\textwidth]{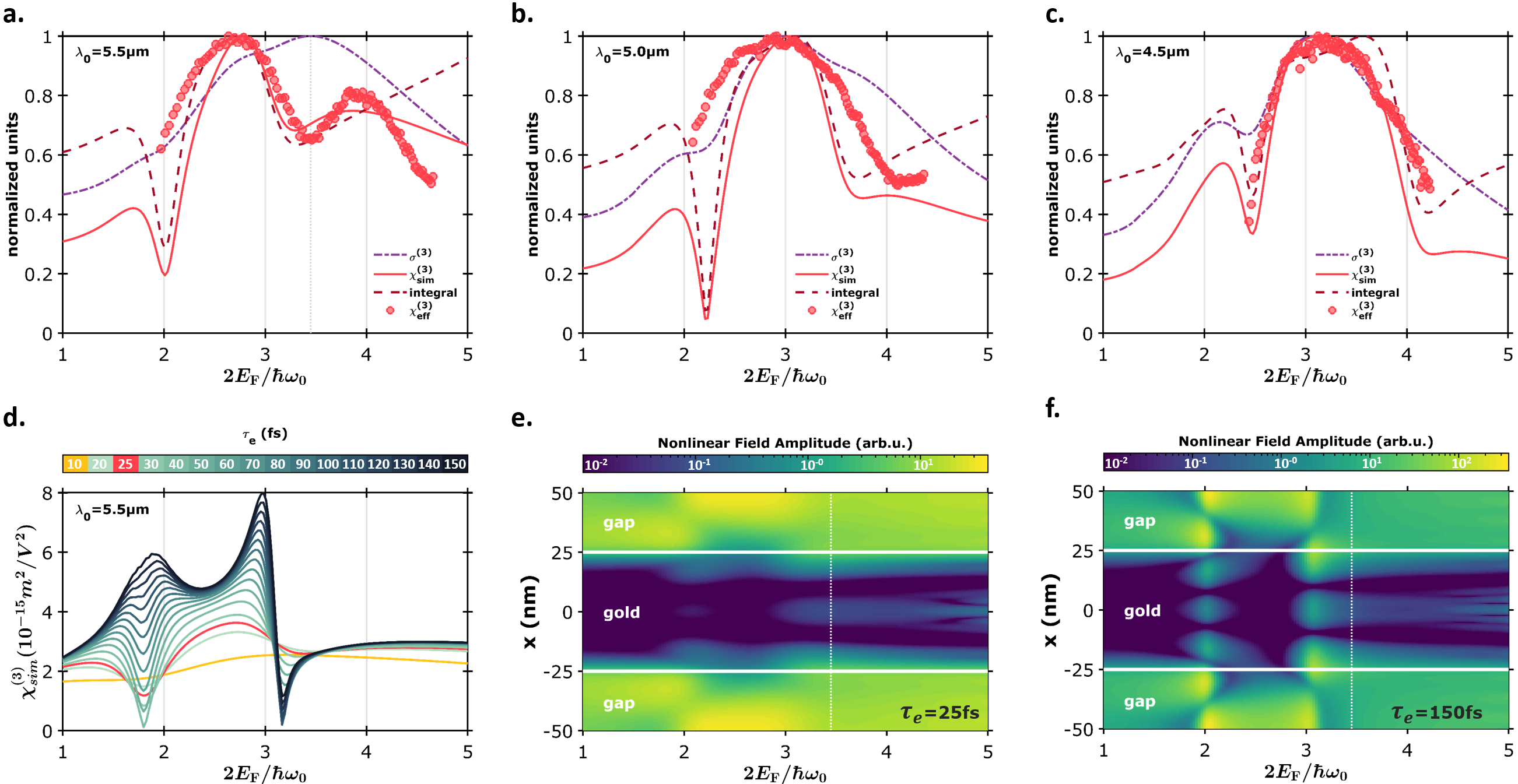}
\caption{\textbf{The role of graphene plasmons in third-harmonic generation.}
\textbf{a-c.} 
Gate dependence of the nonlinear response functions for incident light wavelengths $\lambda_{0}= 5.5,5.0,4.5\,\mathrm{\mu m}$, respectively with a nanoribbon width of $55~\mathrm{nm}$ and a gap of $45~\mathrm{nm}$. All curves are normalised to their maxima in order to qualitatively compare their features. Data in these three panels correspond to the measured $\chi_{\mathrm{eff}}^{(3)}$ (red dots) and simulated $\chi_{\mathrm{sim}}^{(3)}$ (solid-red curve) third-order nonlinear susceptibility, the third-order nonlinear conductivity $\sigma_{\mathrm{sim}}^{(3)}$ (dashed-dotted violet curve), and the nonlinear in-plane field integral (dashed-fuchsia curve).
As the incident photon energy is increased (wavelength decreased) the dip associated with the excitation of an acoustic graphene plasmon moves to higher values of $\EF$.
\textbf{d.} Simulated gate dependence of the third-order nonlinear susceptibility $\chi_{\mathrm{sim}}^{(3)}$ for various electron relaxation times $\tau_{\mathrm{e}}$.
For very low $\tau_{\mathrm{e}}$ no plasmonic effects are predicted by our model, whereas for larger $\tau_{\mathrm{e}}$ plasmonic effects become evident.  Our measurements are explained well by an intermediate value of $\tau_{\mathrm{e}} = 25$ fs (red curve).
\textbf{e.}, \textbf{f.} Product of field amplitudes $|E_x(\omega)^3E_x(3\omega)/E_0|$ in the graphene layer below the gold nanoribbons and in the gap as a function of $\EF$ for $\lambda_{0}=5.5 \mathrm{\mu m}$ and plasmon lifetimes $\tau_{\mathrm{e}}=25\,\mathrm{fs}$ and $\tau_{\mathrm{e}}=150\,\mathrm{fs}$, respectively. An acoustic plasmon can be observed below the gold nanoribbon at $3.45\times \mathrm2{\EF/\hbar\omega_0}$. Because the field of the acoustic plasmon has a sign opposite to the field in the gap (see phase plots in the Appendix, Fig. S6) the net integrated field decreases, leading to a dip in the effective nonlinearity ($\chi_{\mathrm{exp}}^{(3)}$ and $\chi_{\mathrm{sim}}^{(3)}$).
Simulations in panels a-f were carried out with the corresponding electron temperatures specified Fig. S7 and with a nanoribbon width and gap of $50~\mathrm{nm}$.}
\label{img:Figure4}
\end{figure*}

The observed THG signal in bare graphene is plotted in Fig. \ref{img:Figure3}b as a function of $\EF$ {for four different incident wavelengths $\lambda_{\mathrm{0}}=[5.5,5.0,4.5,4.1]\,\mathrm{\mu m}$ ($[0.225,0.248,0.276,0.302]\,\mathrm{eV}$).}
A prominent peak in the THG is observed that emerges at larger $\EF$ for shorter wavelengths and thus larger $\hbar\omega_0$.   
This is more clearly shown in {Fig. S5d} of the Appendix, where we plot the Fermi energy at which the THG is maximised versus the incident photon energy.
As we show in the Appendix, the exact location of the maximum Fermi energy is affected by the electron temperature, which depends on the incident fluence.
Importantly, these gating data show that we can actively modulate the THG signal in bare graphene.
In particular, over the four data sets for bare graphene presented in Fig. \ref{img:Figure3}b, we achieve an average intensity modulation of the THG by a factor $\approx 10\pm3$. As illustrated in Fig. S5 of the Appendix, the high electron temperature (which we estimate to be $\approx 1100\,\mathrm{K}$ in the Methods Section) is the main limitation on this modulation.

Similar gating measurements on the heterostructure for the geometry in which we obtain a maximum field enhancement ($W=200\,\mathrm{nm}$, $g=50\,\mathrm{nm}$) are shown in Fig. \ref{img:Figure3}c.
Once again, we observe the expected shift as a function of the energy of the incident photons.  We can modulate the THG by a factor of $7.4\pm0.2$ with a $3.9\,\mathrm{\mu m}$ incident light wavelength, and by a factor of $4.7\pm0.2$ with $5.5\,\mathrm{\mu m}$.
We stress that this active tunability comes from the unique combination of atomic thickness and linear electronic dispersion in graphene, which cannot be achieved in standard noble metal plasmonics \cite{Kauranen}.

\section*{Plasmon mediation}
Interestingly, the THG signal for a larger range of $\EF$ and for structures with smaller nanoribbon widths reveals an intriguing gate response.  We plot the THG gate response of a heterostructure with $W=55\,\mathrm{nm}$ and $g=45\,\mathrm{nm}$ for different incident wavelengths in Fig. 4a-c.
When the incident wavelength is $5.5~\mathrm{um}$, two peaks and a dip are clearly present in the data (Fig. 4a).
As expected, when the wavelength (energy) is decreased (increased) these features shift to higher Fermi energies. The dip is still visible with an incident wavelength of $5.0~\mathrm{um}$ at $2\EF/\hbar\omega_0\approx 4$ (Fig. 4b). However, it shifts beyond our accessible gating range at a wavelength of $4.5~\mathrm{um}$ (Fig. 4c).
None of these features are evident in the bare graphene data presented in Fig. 3b.

To provide an interpretation, we note that the graphene third-order nonlinearity associated with THG is determined by the interplay between the third-order nonlinear graphene conductivity $\sigma^{(3)}_{3\omega}$ and the nonlinear field integral $\int_{L}{E_x(\omega)^3}E_x(3\omega)dx$ (\textit{i.e.}, the cubic of the linear field is associated with the THG current amplitude, while the field at $3\omega$ represents the emission enhancement produced by the heterostructure at the THG frequency). Tuning the Fermi level to higher energies can lead to the excitation of acoustic graphene plasmons under the metal \cite{Alcaraz} that also affects the nonlinear response. The third-order conductivity $\sigma^{(3)}_{3\omega}$ depends on the excitation frequency $\omega$, Fermi energy $\EF$, the electron relaxation time $\tau_{\mathrm{e}}$, and electron temperature $T_{\mathrm{e}}$ associated with incident light absorption; the latter quantity can reach high values relative to the ambient room temperature that cause anomalous behaviour, such as shifts in the features of the optical conductivity as a function of doping, compared to constant $T_{\mathrm{e}}$ (see S4 of the Appendix). Similarly, these parameters affect the nonlinear field integral through the linear graphene conductivity $\sigma^{(1)}$, although in a different manner, thus leading to different field dependencies on $\EF$ in the gap and below the metal (see simulations in Fig 4e,f). In particular, the fields from different spatial regions can have opposing signs, thus affecting the overall nonlinear performance.

It is thus important to stress that, as we show in the Fig. 4a-c, the third-order conductivity alone, despite its rich dependence on these parameters, cannot fully explain the observed THG signal. This is more evident in Figure S5 of the Appendix, wherein we plot the third-order conductivity at various electron temperatures, observing no features qualitatively similar to the dips in our experimental data. However, the peaks and dips appearing in the response at specific values of $2\EF/h\nu$ correlate well with the integral of the electric fields, stemming in part from plasmonic interferences, and driving the nonlinear response. This integral, shown as a dashed line in Fig. 4a-c, is found to exhibit the same trend as the data. 

From our simulations, we conclude that the nonlinear response in our experiment is mainly driven by the field in the gap region where it takes comparatively larger values (see Fig \ref{img:Figure4}e). In particular, with an incident wavelength of $\lambda = 5.5\,\mathrm{\mu m}$, $W=55\,\mathrm{nm}$, $g=45\,\mathrm{nm}$, and $s=5\,\mathrm{nm}$, we find a dip between the $3\hbar\omega_0$ and $4\hbar\omega_0$ transitions, and a peak just below the $4 \hbar\omega_0$ transition, at $3.45\times \mathrm2{\EF/\hbar\omega_0}$ and $\sim 3.9\times \mathrm2{\EF/\hbar\omega}$, respectively.
The dip in the data can be explained by the partial cancellation of positive and negative complex components  of the incident field in the gap by the field under the gold nanoribbons, which produces a reduction of the observed $\chi^{(3)}$.

In a more intuitive picture, at the dip we excite acoustic graphene plasmons under the metal ribbons \cite{Alcaraz,lee2019graphene}.
As shown in Fig. \ref{img:Figure4}d, for short lifetimes $\tau_\mathrm{e}<100\,\mathrm{fs}$, exciting an acoustic plasmon decreases the net integrated nonlinear field (dashed maroon curves in panels a-c, labeled `integral'). This manifests as a dip in the nonlinear signal.
In Fig. \ref{img:Figure4}a-c, the model with $\tau_\mathrm{e} =25\,\mathrm{fs}$ fits best to our data, confirming that the dips at $1.8$ and $3.45\times2\EF/\hbar\omega$ in Fig. \ref{img:Figure4}a correspond to the excitation of acoustic plasmons. These plasmonic features do not appear in the $W=200\,\mathrm{nm}$ heterostructure data presented in Fig. \ref{img:Figure3}c because the width is too large, and exciting acoustic plasmons in these structures would require more doping or a longer incident wavelength.
As an additional experimental confirmation, in panel b and c we repeat the same gate-dependent THG measurements for higher incident photon energies $0.25\,\mathrm{eV}$ ($5.0\,\mu m$) and $0.28\,\mathrm{eV}$ ($4.5\,\mu m$). This shifts the acoustic plasmon resonance to higher gate voltages, causing the dips to move to higher Fermi energies in our simulations.
{This shift is a key signature of graphene plasmons, which} we confirm experimentally, observing that the dip disappears for high incident phototon energies.

Finally, when graphene plasmons are explicitly suppressed {in our simulations (}by decreasing the plasmon lifetime $\tau_{\mathrm{e}}$), we are unable to reproduce the dips and peaks in our data (see the purple curve in Figure \ref{img:Figure4}a), further confirming the excitation of plasmons in our graphene heterostructures. 

\section*{Conclusion}
Recent studies have reported a wide range in the estimate of $\chi^{(3)}$ using third-harmonic generation \cite{Jiang, Kumar, Hong, Saynatjoki, Woodward, soavi2018broadband}.
By calculating an effective nonlinearity for the $W=200\,\mathrm{nm}$ heterostructure with a monolayer h-BN spacer,  $\EF=0.45\,\mathrm{eV}$, irradiated with $5.5\,\mathrm{\mu m}$ light, and an electron temperature of approximately $1100\,\mathrm{K}$, we find $\chiexp\approx 3.4 \times 10^{-6}\,{\rm esu}$, which is an order of magnitude larger than that of bare graphene, for which we obtain a maximum value of $\chiexp\approx 3.9 \times 10^{-7}\,{\rm esu}$ (with $\EF=0.39\,\mathrm{eV}$ and irradiated at $5.5\,\mathrm{\mu m}$).
Moreover, our all experimental measurements agree well with simulations based on the third-order nonlinear conductivity taken from References ~\cite{mikhailov2016quantum,Rostami}.

Unlike in metal plasmonics, we can actively modulate the nonlinearity of our graphene heterostructures by controlling $\EF$ with an external gate voltage. Graphene-based linear optical devices have already been shown to operate at GHz {frequencies} \cite{Phare} and hence, our system provides a new route toward {high-speed} nonlinear optoelectronic switches and frequency converters.
Additionally, our measurements reveal intriguing plasmonic effects supported by simulations in which graphene surface plasmon polaritons appear to directly modify the nonlinear optical response of our structures. These plasmonic excitations potentially provide a novel approach to the manipulation and amplification of light at subwavelength scales. In the present work, acoustic graphene plasmons seem to modulate the nonlinear response of graphene, while our simulations suggest that improving the plasmon lifetime by a factor of five would already increase the nonlinear response by one order of magnitude. We also observed that the maximum field enhancement (and hence the maximum nonlinearity) was obtained in a device that did not support acoustic graphene plasmons.
However, one could in principle design different metal nanostructure geometries to simultaneously enhance the field and launch graphene plasmons. Engineering smaller nanostructures also has the potential to excite plasmons at shorter wavelengths. Our findings suggest that graphene plasmonic devices could provide unprecedentedly strong nonlinearities, potentially resulting in nonlinear optical effects at the single-photon level \cite{tame2013quantum,Gullans,Calafell}.  

\section*{acknowledgments}
The authors thank Simone Zanotto for assistance with the Matlab code.
P.W. acknowledges support from the European Commission through ErBeSta (No. 800942), the Austrian Research Promotion Agency (FFG) through the QuantERA ERA-NET Cofund project HiPhoP,  from the Austrian Science Fund (FWF) through CoQuS (W 1210-N25), NaMuG (P30067-N36) and BeyondC (F 7113-N38), the U.S. Air Force Office of Scientific Research (FA2386-233 17-1-4011 and FA8655-20-1-7030) and Red Bull GmbH. 
F.J.G.A. acknowledges support from the ERC (Advanced Grant 789104-eNANO) and the Spanish MINECO (MAT2017-88492-R). 
ICFO is finantially supported by the Spanish MINECO (SEV2015-0522), the Catalan CERCA, Fundaci\'o Privada Cellex, the Spanish Ministry of Economy and Competitiveness, through the "Severo Ochoa" Programme for Centres of Excellence in R$\&$D (SEV-2015-0522) and Fundacio Cellex Barcelona, Generalitat de Catalunya through the CERCA program.
F.H.L.K. acknowledges  support from the Government of Spain (FIS2016-81044; Severo Ochoa CEX2019-000910-S), Fundacio Cellex, Fundacio Mir-Puig, and Generalitat de Catalunya (CERCA, AGAUR, SGR 1656). Furthermore, the research leading to these results has received funding from the European Union's Horizon 2020  under grant agreements no.785219 (Graphene flagship Core2) and no. 881603 (Graphene flagship Core3). This work was supported by the ERC TOPONANOP under grant agreement n 726001.
The MIT portion of this work was supported in part by the NSF Center for Integrated Quantum Materials (CIQM), the US Army Research Office (Award W911NF-17-1-0435), and the Institute for Soldier Nanotechnologies (contract number W911NF-18-2-0048).
The Center for Nano Optics is financially supported by the University of Southern Denmark (SDU 2020 funding). J.D.C was supported by VILLUM Fonden (grant No. 16498).
I.A.C., P.K.J. acknowledges support from the University of Vienna via the Vienna Doctoral School. 
L.A.R. acknowledges support from the Templeton World Charity Foundation (fellowship no. TWCF0194).
D.A.I. acknowledges support from the Spanish MINECO FPI Grant (BES‐2014‐068504).   
A.T. acknowledges support from the European Union’s Horizon 2020 research and  innovation programme under the Marie Skłodowska‐Curie grant agreement No 801110 and the  Austrian Federal Ministry of Education, Science and Research (BMBWF).
It reflects only the author's view, the EU Agency is not responsible for any use that may be made of  the information it contains.

\section*{Methods}
\noindent\textbf{Experimental details---}
We carry out our THG measurements using a modified z-scan setup, where the TH signal is measured while the sample is moved along the \textit{z} axis through the focus of the laser beam (see Fig. \ref{img:Figure1}d). Our incident light beam consists of linearly-polarised pulses of $\sim260$\,fs duration with a tunable carrier wavelength of $5.5\,\mathrm{\mu m}$ ($0.225$\,eV) at a 80\,MHz repetition rate, which we create by performing Difference-Frequency Generation (DFG) between the signal and the idler beams of an Optical Parametric Oscillator (OPO). The typical acquisition time for a gate-dependent THG measurement is around $30$\,min; over this time we observe $<2\,\%$ power fluctuations. We use a Half-Wave Plate (HWP) to tune the polarisation of the incoming beam to that set by the polariser (pol). By rotating the HWP and the polariser together we can rotate the incident light polarisation relative to the orientation of the nanoribbons.
Then, a lens with a $5.26$\,mm focal length focuses the incident light down to a waist of $\approx 20\,\mathrm{\mu m}$ for $5.5\,\mathrm{\mu m}$ nm light and  $\approx 13\,\mathrm{\mu m}$ for $3.9\,\mathrm{\mu m}$ light. When the sample is moved parallel to the incident light beam (along the $z$ axis), the nonlinear emission occurs most efficiently where the fluence is maximum (at the focal point). 
Afterwards, a lens with a $11\,\mathrm{mm}$ focal length collimates the beam, which is then sent through a bandpass filter (BP) to separate the THG signal from the excitation beam. 
Finally, the signal is coupled into a multimode fiber and sent to a large-area Superconducting Nanowire Single-Photon (SNSP) detector, with a $\approx 20\,\%$ detection efficiency at the third-harmonic wavelength, $5.5/3\,\mathrm{\mu m} = 1.833\,\mathrm{\mu m}$.

We verify the wavelength of this signal by removing all spectral filters, keeping the sample in the focus, and using a NIREOS GEMINI interferometer to perform Fourier transform spectroscopy on the signal.

\noindent\textbf{Sample fabrication and electrical doping---}
Here we use a heterostructure consisting of wet-transferred chemical-vapor deposition (CVD) graphene stripes shaped by dry ion etching (RIE) with a UV resist positive mask. The standard CVD graphene used in these samples has typical mobility below $1000\,\mathrm{m^2/Vs}$ and $\approx25\,\mathrm{fs}$ plasmon lifetimes, which is much shorter than those found in exfoliated graphene, where it can reach up to $\approx500\,\mathrm{fs}$ at room temperature \cite{Woessner}. An insulator, which can consist of either wet-transferred CVD h-BN or $\mathrm{Al_2 O_3}$ deposited by atomic layer deposition (ALD), defines the space between the substrate and the nanometer-thick gold nanoribbons, which are deposited by thermal evaporation onto a positive polymethyl methacrylate (PMMA) mask shaped using e-beam lithography. Note that for adhesive purposes, there is a $2\,\mathrm{nm}$-thick Ti layer below the gold nanoribbons. These nanoribbons concentrate the electric field of a far-field incident light into graphene \cite{Alcaraz}.

We perform the gating measurements applying a backgate voltage with a sourcemeter that allows us to monitor the current leakage between the Si and graphene layers, while increasing the applied voltage. A $1\,\mathrm{mV}$ voltage between the source and drain allows us to measure the graphene resistance, which we then use to estimate the induced $\EF$. In order to calculate the induced $\EF$ given an applied voltage $V$, we consider the $\mathrm{SiO_2}$ and graphene layers to behave as a parallel-plate capacitor, with the $\mathrm{SiO_2}$ dielectric with relative permittivity $\epsilon_d\approx 4$ and thickness $d_d$ in between. The conversion is given by
\begin{equation}
\EF(V)=sign\{n\}\hbar\sqrt{\pi\left|n\right|}
\end{equation}
with $n=C\cdot\Delta V/e$, where $C=\epsilon_0\epsilon_d/d_d$ is the surface capacitance and $\Delta V=V-V_{\mathrm{CNP}}$ is the difference between the applied voltage and the voltage at which the charge neutrality points (CNP) is found.

\noindent\textbf{Extracting the third-order susceptibility---} Experimentally we extimate $\chi_{exp}^{(3)}$ starting with the expression of the input ($i$) and output ($o$) average power as a function of the field as,
\begin{equation}
P(\omega_{i,o})=\frac{1}{8}\left(\frac{\pi}{\ln{2}}\right)^{3/2}f\tau W^2 n_{\omega_{i,o}}\epsilon_0 c \frac{|E(\omega_{i,o})|^2}{2},
\end{equation} 
where we assume laser pulses with repetition rate $f$, duration $\tau$, waist $W$ on the sample and Gaussian profile, and $n_{\omega_{i,o}}=2.4$ is the refraction index, and $\epsilon_0$ and $c$ are the permittivity and speed of light in vacuum \cite{Jiang}.
Additionally, we can write the THG process as a function of the input and output fields as follows and solve for $\chi_{\mathrm{exp}}^{(3)}$:
\begin{equation}
E(\omega_{\mathrm{o}})=\frac{1}{4}\frac{\ii\omega_{\mathrm{i}}}{2\pi c}\chi_{\mathrm{exp}}^{(3)} \dgr E^3(\omega_{\mathrm{i}}),
\end{equation}
where $\omega_{\mathrm{i}}=\omega_{\mathrm{o}}/3$ and $\dgr=0.33\,\mathrm{nm}$ is the effective thickness of graphene.

\noindent\textbf{Electron temperature---}
In order to estimate the electron temperature in our samples, we have performed gating measurements on bare graphene for different incident wavelengths (see Fig. \ref{img:Figure3}b) and determined the Fermi energy at which the maximum third-order susceptibility is found. As shown in Fig. S5d, this Fermi energy at which the maximum $\chi^{(3)}$ is found shifts linearly with the photon energy of the excitation light. We fit the electron temperature of the simulations to best match the experimental data and determine a value of $1100$\,K, which is in good agreement with the simulations shown in Fig. S7. Note that, as mentioned in Ref. \cite{Soavihot, Soavi, Jiang, Mikhailov19}, the electron temperature changes with both the excitation wavelength and the Fermi energy. However, due to the small Fermi energy shift of about $80$\,meV, this effect was neglected in our simulations.

For the power dependent measurements, it is necessary to describe the electron temperature as a function of fluence. To a good approximation, we assume a chemical potential $\mu > k_{\mathrm{B}} T_{\mathrm{e}}$, which allows us to describe the electron temperature as follows \cite{Shi14}: 
\begin{equation}
T_e = \sqrt{1 + \frac{2 \gamma F}{\alpha T_0^2}}, \qquad \text{where } \alpha = \frac{2\pi}{3}\frac{k_{\mathrm{B}}^2 \mu}{\hbar^2 v_{\mathrm{F_0}}^2},
\end{equation}
$T_0$ is the ambient temperature, $F_0$ is the energy per pulse of the incident light and $\gamma$ is the amount of energy absorbed that leads to hot electrons, which we have considered to be $3.5\times 10^{-3}$, which is consistent with previously reported values\cite{Shi14}.

Considering the different electron temperatures for different excitation powers as discussed in Ref.\cite{Soavihot}, we are able to explain the power dependence of the TH signal (see Fig. \ref{img:Figure2}b).

\noindent\textbf{Simulating the third-order susceptibility---} 
We write the third-harmonic susceptibility as
\begin{equation}
\chisim=\left|\frac{\alpha_{3\omega}^{(3)}}{\epsilon_{0}\dgr L}\right|,
\label{eq:chi3_sim}
\end{equation}
where $\alpha_{3\omega}^{(3)}$ is third-order polarisability given by

\begin{equation}
\alpha_{3\omega}^{(3)}(\omega)=\frac{\ii\sigma_{3\omega}^{(3)}}{3\omega}\int_{L}{\eta_{\omega}^3(x)}\eta_{3\omega}(x)dx,
\label{eq:alpha_sim}
\end{equation}
where $\sigma_{3\omega}^{(3)}$ is the analytical third-order conductivity in graphene derived by Mikhailov in Ref.\cite{mikhailov2016quantum}, and $L$ is the length of the simulated region used for integration along the direction perpendicular to the nanoribbon width. Here, $\eta_{\omega}^3(x)$ is a dimensionless quantity representing the enhancement in electric field amplitude acting on the graphene layer relative to the incident field amplitude, so when multiplied by $E^3(\omega_i)$ in Equation (4) it yields the actual field amplitude as a function of position $x$; likewise, $\eta_{3\omega}(x)$ represents the factor by which the THG field reaching the detector is modified by the presence of the structure surrounding the graphene. Equation (7) represents the contribution of the THG current to the far field. In particular, the $\eta_{\omega}^3(x)$ factor times $\sigma_{3\omega}^{(3)}$ is the THG current, which we represent as a polarisation density at the emission frequency $3\omega$ through the continuity equation. Because of reciprocity, $\eta_{3\omega}(x)$ is exactly given by the enhancement amplitude in the near field relative to free space when the structure is illuminated with $3\omega$ radiation. More precisely, we calculate $\eta_{3\omega}(x)$ as the complex factor of enhancement relative to the incident field in the field acting on the graphene layer when it is illuminated by a $3\omega$ plane wave impinging from the detector direction.
The electric field enhancements $\eta_{\omega}(x)$ and $\eta_{3\omega}(x)$ entering the above expression are obtained using a RCWA 
Matlab script \cite{Manceau} and adapted to include graphene as an interface material adopting the nonlocal 2D linear optical conductivity of graphene $\sigma(Q,\omega)$\cite{Koppens} that depends on the chemical potential $\mu$ and electronic temperature $T_{\mathrm{e}}$. We note that these enhancement factors depend on both the geometry of the heterostructures and the $T_{\mathrm{e}}$ - and $\mu$-dependent linear conductivity.
It requires periodicity of the structure perpendicular to the layered dimension.
Here we consider the influence of electronic heating by the incident light pulse in the optical response of the graphene-metal hybrid system. In particular, following the procedure in Ref.\cite{Yu}, we make use of the implicit relation between $\EF$, $T_{\mathrm{e}}$, and $\mu$ obtained from conservation of doping charge,
\begin{align}
\begin{split}
\left(\frac{\EF}{k_\mathrm{B} T_{\mathrm{e}}}\right)^2=2\int^\infty_0 dx x\Bigl[\left(e^{x-\mu/k_\mathrm{B} T_{\mathrm{e}}}+1\right)^{-1}\\-\left(e^{x+\mu/k_\mathrm{B} T_{\mathrm{e}}}+1\right)^{-1}\Bigl],
\end{split}
\end{align}
along with the graphene heat capacity
\begin{equation}
F=\beta\frac{(k_{\rm B}T)^3}{(\hbar v_{\mathrm{F}})^3},
\end{equation}
where $F$ is the energy of the excitation pulse absorbed (i.e., $F=\eta_a F_0$ where $\eta_a$ is the absorbed fraction of pulse energy $F_0$). The absorbed energy into graphene was obtained from the linear RCWA simulations at an ambient temperature of $300$\,K and
\begin{equation}
\begin{split}
\beta=\frac{2}{\pi}\Biggl\{\int^\infty_0 dx x^2\Bigl[\left(e^{x+\mu/k_{\mathrm{B}} T_{\mathrm{e}}}+1\right)^{-1}\\+\left(e^{x-\mu/k_{\mathrm{B}} T_{\mathrm{e}}}+1\right)^{-1}\Bigl]-\frac{1}{3}\left(\frac{\EF}{k_{\mathrm{B}} T_{\mathrm{e}}}\right)^3\Biggl\},
\end{split}
\end{equation}

The linear conductivity can thus be determined directly and the temperature dependent nonlinear conductivity can be compute by using Maldague’s identity:

	\begin{equation}
		\sigma_{3\omega}^{(3)}(\omega,\tau,\mu,T)=\frac{1}{4k_{\mathrm{B}}T}\int_{-\infty}^{\infty}dE\frac{\sigma_{3\omega}^{(3)}(\omega,\tau,\mu,0)}{\cosh^2{(\frac{E-\mu}{2k_{\mathrm{B}}T})}}.
		\label{eq:T_integral}
	\end{equation}

\bibliography{THG_heterostructures.bbl}
\clearpage
\onecolumngrid
\beginsupplement
\section*{Appendix}

\subsection{Additional Figures}
\begin{figure*}[hpbt]
\centering
\includegraphics[width=0.5\textwidth]{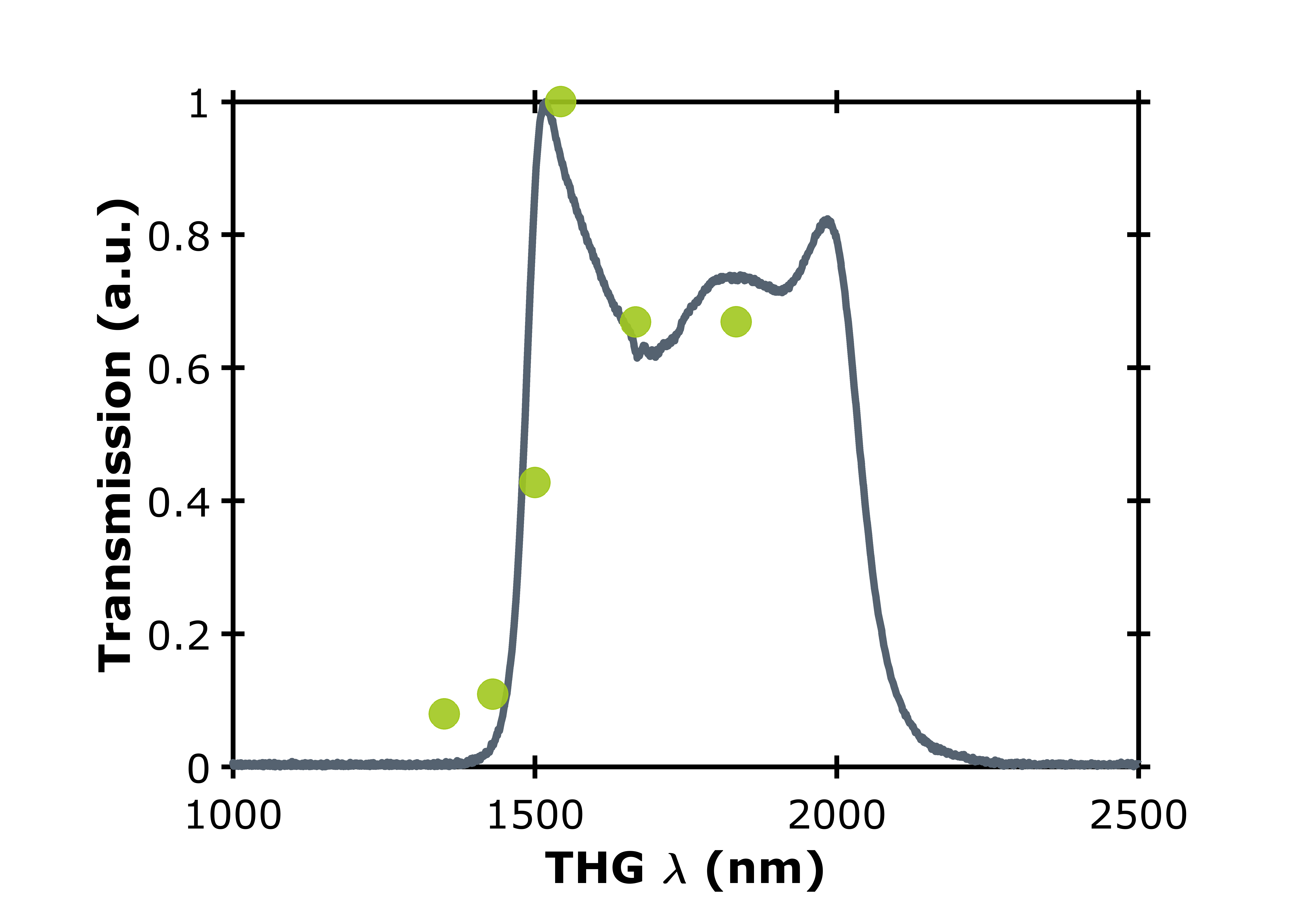}
\caption{\textbf{Confirmation of the third-harmonic signal wavelength with a $\mathbf{5500}$ nm incident beam.}
The transmission of the third-harmonic signal through a bandpass filter at ${1750\,\mathrm{{nm}}}$ with a ${500\,\mathrm{{nm}}}$ bandwidth as the incident wavelength is changed. The points are the meassured transmission of the third-harmonic signal and the solid curve is the transmission of the filter measured with standard FTIR.}
\label{img:FigureS1}
\end{figure*}

\newpage
\begin{figure*}[hpbt]
\centering
\includegraphics[width=\textwidth]{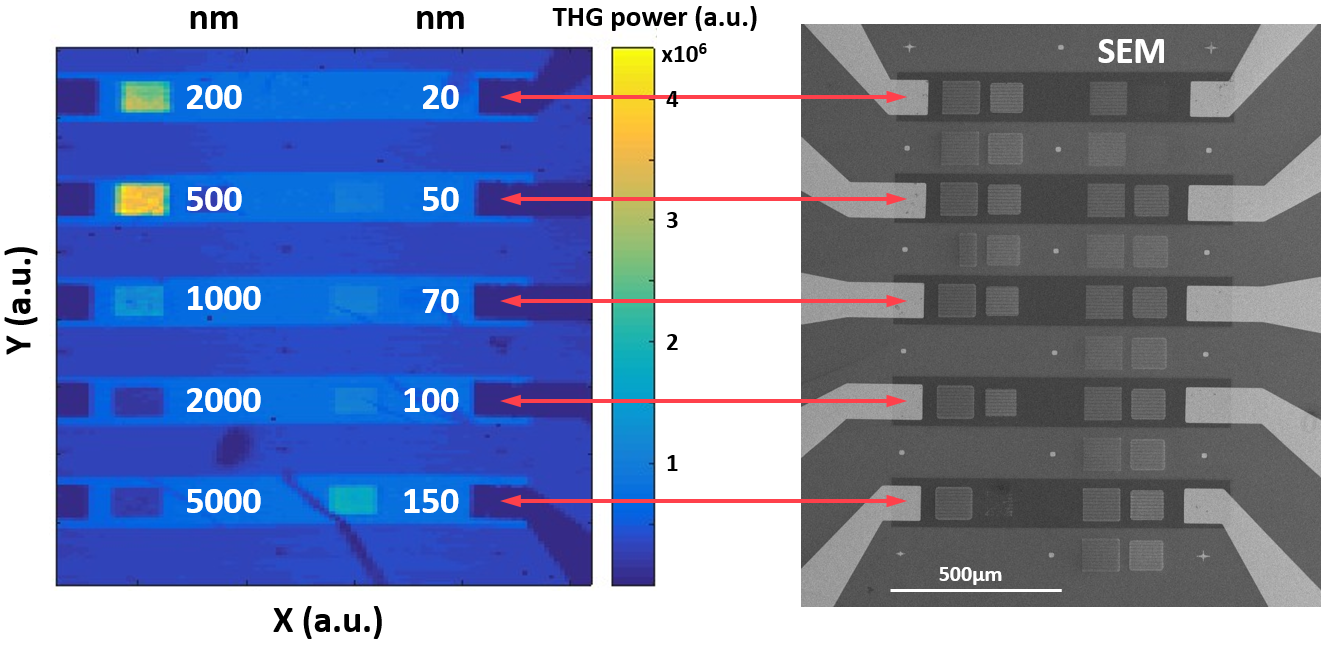}
\caption{\textbf{Comparison of the third-harmonic signal with an SEM image of the sample.}
The left panel shows the third-harmonic signal as the entire sample is moved transversely through the incident beam. The corresponding SEM image of the sample is shown on the right. 
Strong third-harmonic signals can be distinguished in the regions with both graphene and golden nanoribbons, as well as a weaker signals from the strips of bare graphene surrounding the nanoribbons.
There is no discernible signal from the nanoribbons without graphene.
The darker areas indicating the contacts are also evident. 
As a reference, the SEM image on the right hand side confirms the structure of the sample, where the dark horizontal bands indicate the presence of bare graphene. The brighter squares on top of these bands are different arrays of golden nanoribbons with widths $W$ ranging from  $20-5000\,\mathrm{nm}$. Note that some of the arrays noticeable on the SEM image do not show any THG signal. This is due to experimental imperfections of the golden nanoribbons during the fabrication process.}
\label{img:FigureS2}
\end{figure*}

\newpage
\begin{figure*}[hpbt]
\centering
\includegraphics[width=\textwidth]{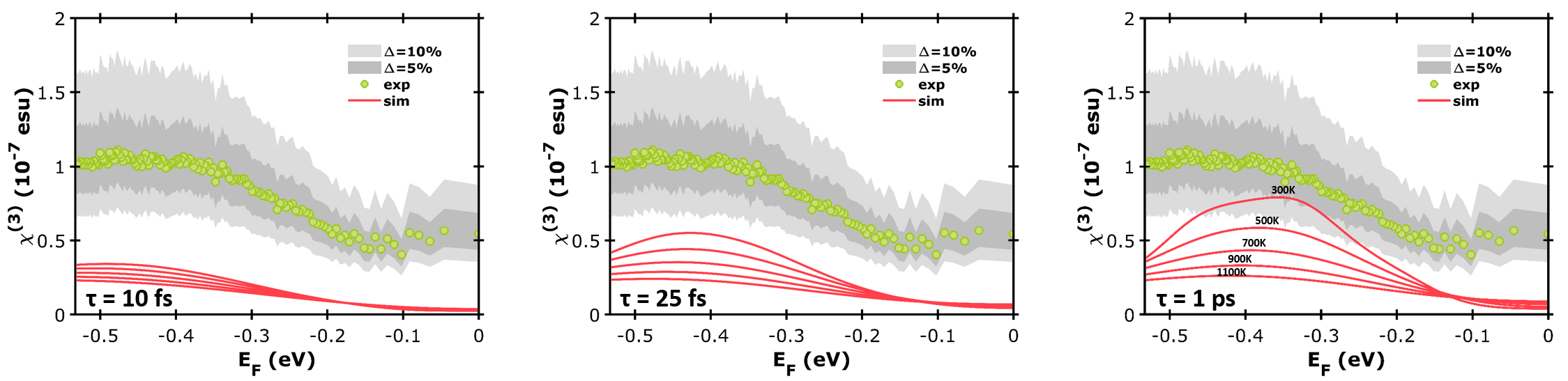}
\caption{\textbf{Comparison of the magnitude of the experimentally estimated $\mathbf{\chi^{(3)}}$ with the calculated value.}
The shaded areas around the green points represent the experimental uncertainty in our estimation of $\chi^{(3)}$. 
We calculate the error bars of the experimental data (green dots) with an uncertainty of $5\%$ and $10\%$ on the experimentally measured parameters, defined by the dark and light grey areas, respectively. The red solid lines show the theoretically calculated $\chi^{(3)}$ at different electron temperatures for $10\,\mathrm{fs}$, $25\,\mathrm{fs}$ and $1\,\mathrm{ps}$, from left to right.}
\label{img:FigureS3}
\end{figure*}

\newpage
\begin{figure*}[hpbt]
\centering
\includegraphics[width=\textwidth]{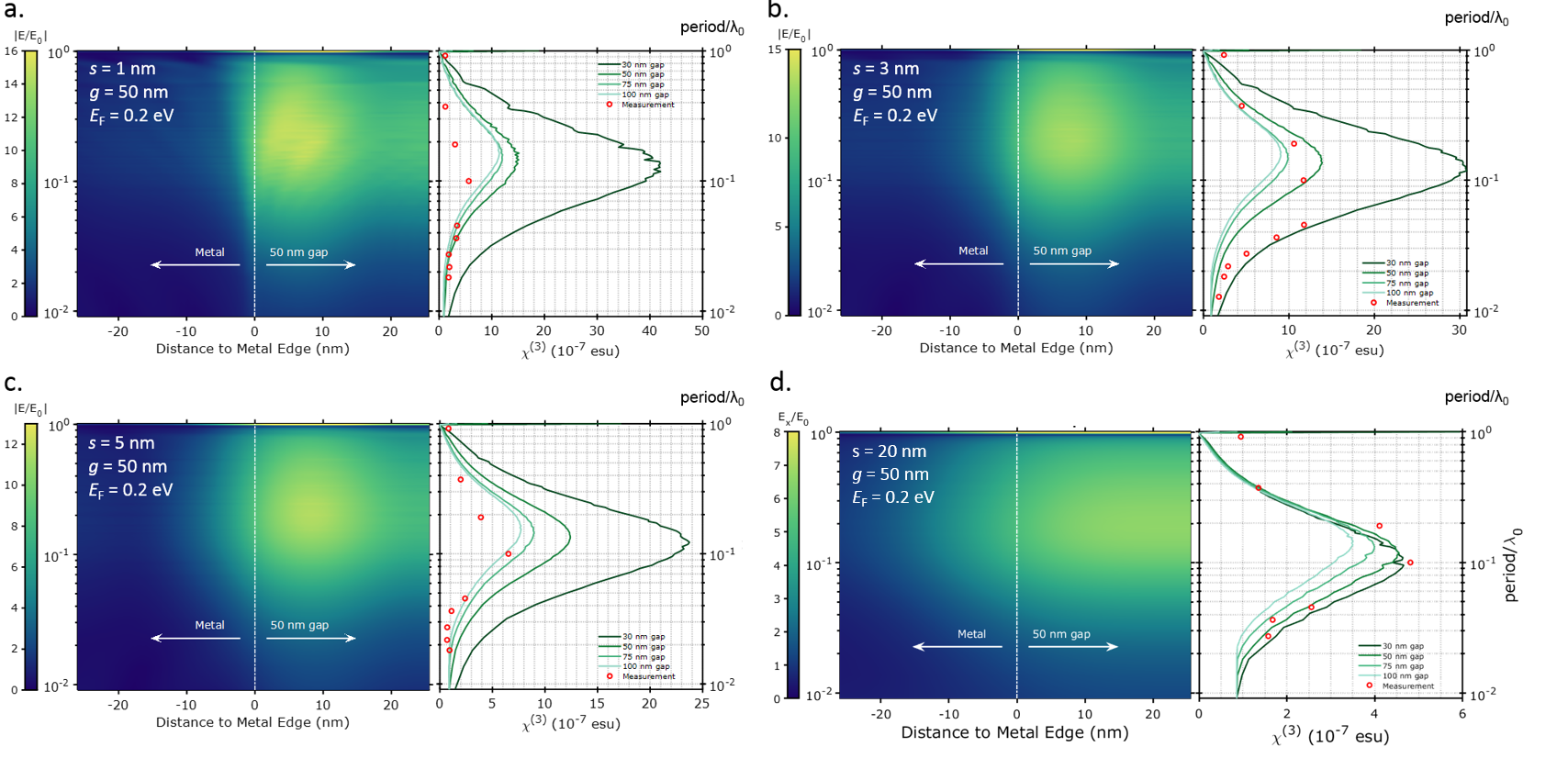}
\caption{\textbf{Field confinement in the gap between the nanoribbons and $\mathbf{\chi^{(3)}}$ as a function of the nanoribbon width.} Each panel corresponds to a different $\mathrm{Al_{2}O_{3}}$ spacing $s = [1,3,5,20]\,\mathrm{nm}$.
}
\label{img:FigureS4}
\end{figure*}

\newpage
\begin{figure*}[hpbt]
\centering
\includegraphics[width=\textwidth]{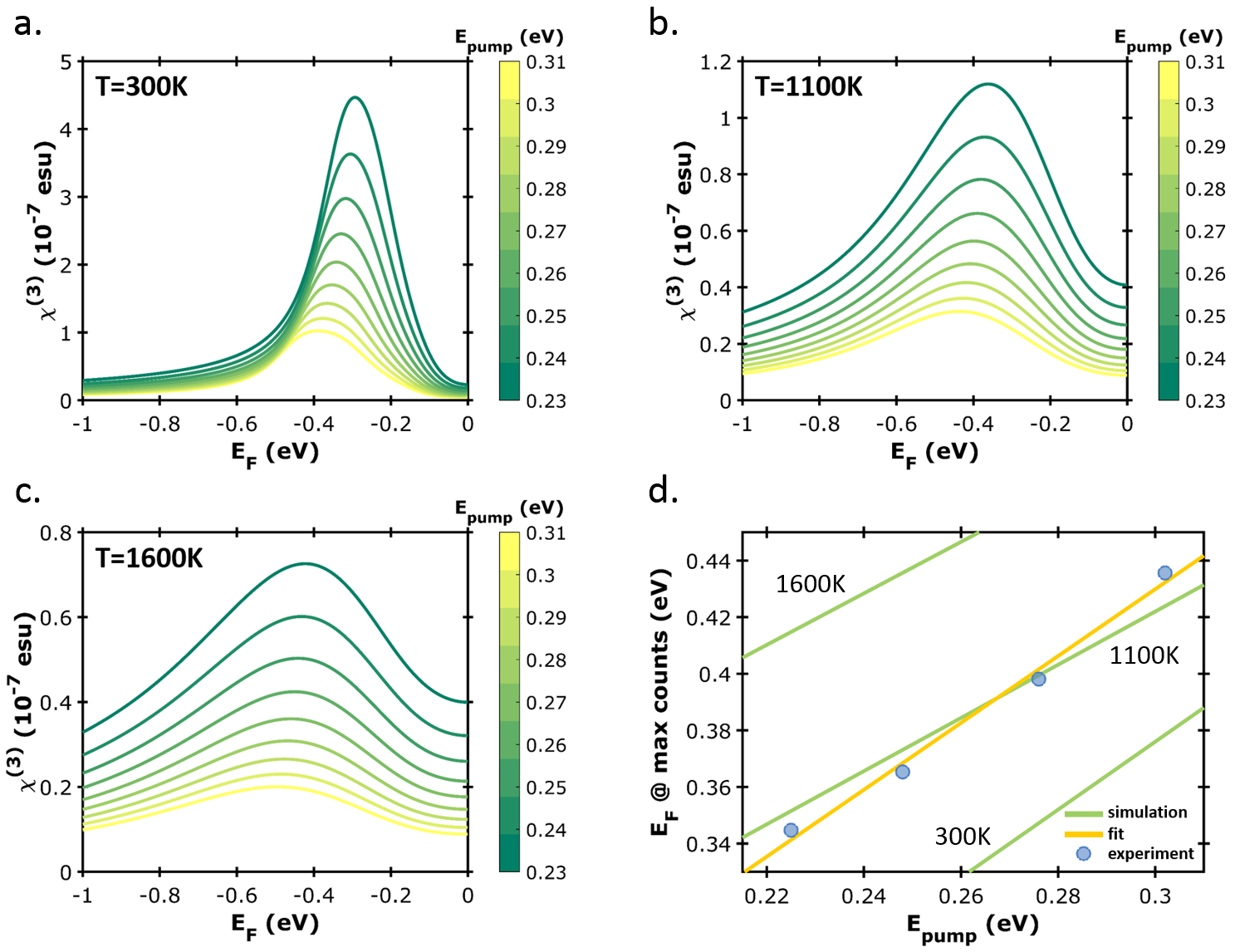}
\caption{\textbf{Effect of the electron temperature on the optimum Fermi energy.} \textbf{a-c.} Simulations of the third-order susceptibility $\chi^{(3)}$ as a function of the Fermi energy $\EF$ for incident wavelengths in the $\lambda_{0}=4.1-5.5\,\mathrm{\mu m}$ ($E_{0}=0.225-0.302\,\mathrm{eV}$) range and electron temperatures of $300\,\mathrm{K}$, $1100\,\mathrm{K}$ and $1600\,\mathrm{K}$, respectively. \textbf{d.} Blue dots show the experimental values of the $\EF$ at which the maximum $\chi^{(3)}$ are found in the gate measurement, for $\lambda_{0}=[5.5,5.0,4.5,4.1]\,\mathrm{\mu m}$ ($E_{0}=[0.225,0.248,0.276,0.302]\,\mathrm{eV}$). The green solid lines are simulations at $T_{\mathrm{e}}=[300,1100,1600]\,\mathrm{K}$ and the solid yellow line is a linear fit of the data points.}
\label{img:FigureS5}
\end{figure*}

\newpage
\begin{figure*}[hpbt]
\centering
\includegraphics[width=\textwidth]{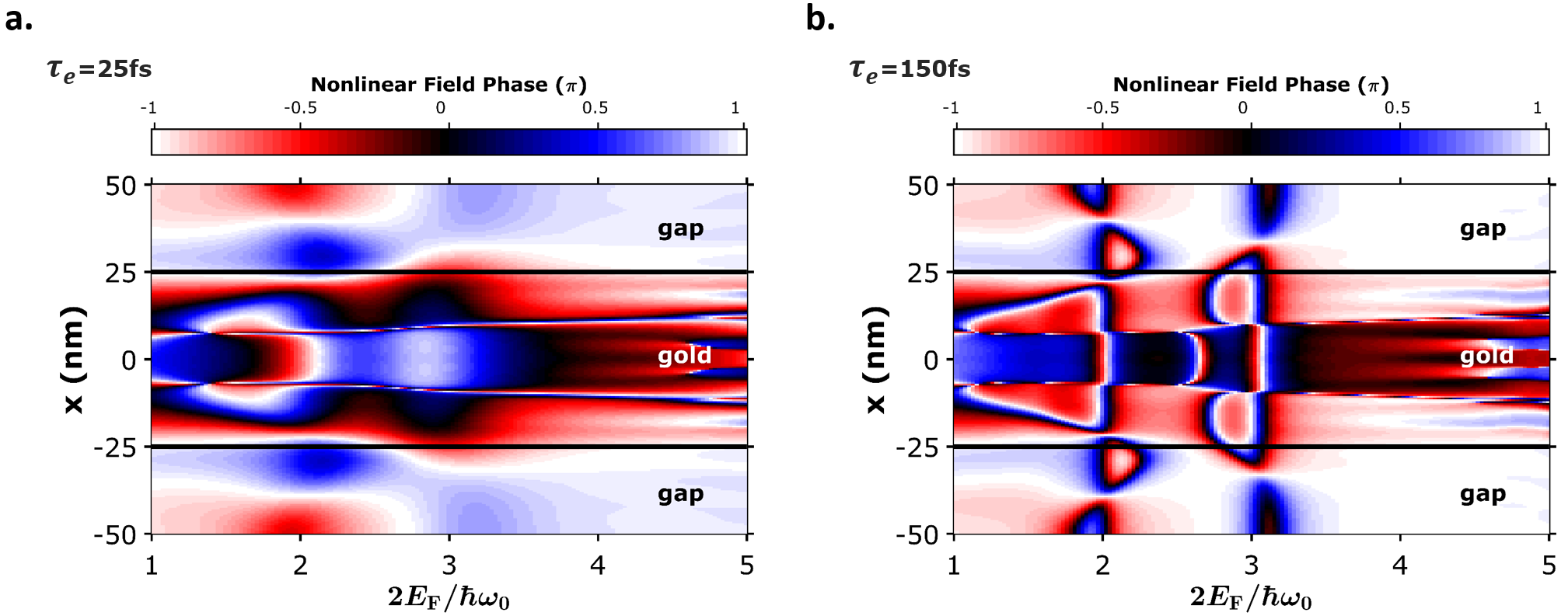}
\caption{\textbf{Phase of the electric field in graphene.} Phase of the simulated $|E_x(\omega)^3E_x(3\omega)/E_0|$ field in the graphene layer below the gold nanoribbons and in the gap, as a function of $\EF$ periods for $\lambda_{0}=5.5 \mathrm{\mu m}$, and plasmon lifetime of \textbf{a.} $\tau_{\mathrm{e}}=25\,\mathrm{fs}$, \textbf{b.} $\tau_{\mathrm{e}}=150\,\mathrm{fs}$.}
\label{img:FigureS7}
\end{figure*}

\newpage
\begin{figure*}[hpbt]
\centering
\includegraphics[width=\textwidth]{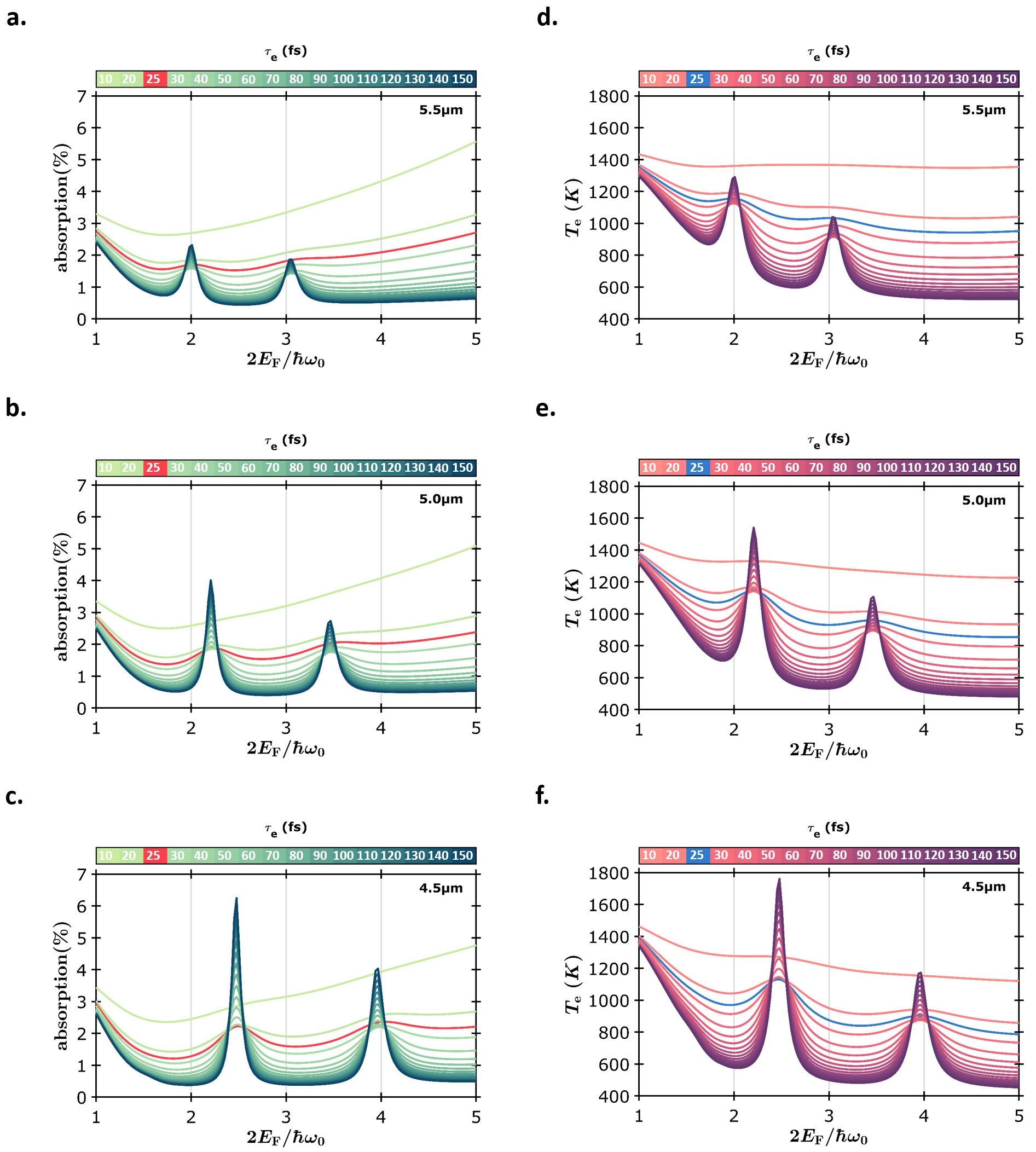}
\caption{\textbf{Light absorption and electron temperature.} \textbf{a.-c.} Absorption of the excitation field in graphene for incident wavelengths $5.5\,\mathrm{\mu m}$, $5.0\,\mathrm{\mu m}$ and $4.5\,\mathrm{\mu m}$, respectively. \textbf{d.-f.} Electron temperature $T_{\mathrm{e}}$ derived from the absorbed excitation field in graphene for incident wavelengths $5.5\,\mathrm{\mu m}$, $5.0\,\mathrm{\mu m}$ and $4.5\,\mathrm{\mu m}$, respectively.}
\label{img:FigureS8}
\end{figure*}

\clearpage
\subsection{Rigorous coupled-wave analysis (RCWA)}
The simulations in Fig. 3a and 4 of the main text are achieved with the Matlab script from Ref.\cite{matlabperiodic,Manceau}. The geometry of the 2D simulation is given by the layer stack of the original chip, which is composed by a $525$\,$\mu$m-thick silicon chip coated with a $285$\,nm-thick SiO$_2$ layer above and below. The monolayer graphene placed on the front side of the chip is modelled by a surface nonlocal conductivity, given in Ref.\cite{Koppens}. The spacer layer built by h-BN or Al$_2$O$_3$ with thickness $s$ was placed on top of the graphene. Finally a $10$\,nm-thick gold nanoribbon with width $W$ was placed on top of the spacer. In a good approximation, the electric permittivity for silicon was assumed to be $12$ and wavelength independent. The wavelengths depending dielectric permittivity of SiO$_2$, h-BN and gold are found in Ref.\cite{Kischkat2012}, Ref.\cite{Caldwell2014} and Ref.\cite{Olmon2012}, respectively. As mentioned in the Methods, the simulation script requires a periodic structure in the lateral dimension. Here the periodicity is given by the sum of the nanoribbon width and the gap in-between. The electric field of the excitation incident light (THG signal) in the graphene layer was calculated for a normal incident of a monochromatic plane wave polarized parallel to the $x$ axis and angular frequency $\omega$ ($3\omega$). From the electric fields $E_x(\omega)$ and $E_x(3\omega)$, $|\chi^{(3)}_\mathrm{sim}|$ was calculated as described in the Methods section.\\
For accessing the electron temperature $T_{\mathrm{e}}$ in the graphene layer, we calculate the amount of absorbed energy in the graphene. To do so, we run the simulations with the RCWA, considering an ambient temperature of $300\,\mathrm{K}$ (see Fig. S7a-c.
Solving the implicit Eq. $8-10$ of the main text leads to the electron temperature used in Fig. 4. These temperatures were considered in the third-order nonlinear conductivity and are shown in Fig. S7d-f.
During the gating measurements, when the voltage is applied between the graphene and silicon substrate, a leakage current through the SiO$_2$ substrate can occur. Because of that, the charge densities in the graphene layer may differ slightly from those determined via the capacitor model introduced in the Methods section. This difference depends on the measurement time and the leakage. In order to compensate for this effect, the gating voltage of the experimental data shown in Fig. 4a-c of the main text has been fitted accordingly.

\end{document}